\documentclass{osa-article}

\journal{osajournal}

\articletype{Research Article}

\begin{document}

\title{Highly angular resolving beam separator based on total internal reflection}

\author{Moritz Mihm,\authormark{1,*} Ortwin Hellmig,\authormark{2} Andr{\'{e}} Wenzlawski,\authormark{1} Klaus Sengstock,\authormark{2} and Patrick Windpassinger\authormark{1}}

\address{\authormark{1}Johannes Gutenberg-Universit{\"{a}}t Mainz, Staudingerweg 7, 55128 Mainz, Germany\\
\authormark{2}Institut f{\"{u}}r Laserphysik/Zentrum f{\"{u}}r optische Quantentechnologien, Universit{\"{a}}t Hamburg, Luruper Chaussee 149, 22761 Hamburg, Germany}

\email{\authormark{*}mmihm@uni-mainz.de} 



\begin{abstract}
We present an optical element for the separation of superimposed beams which only differ in angle. The beams are angularly resolved and separated by total internal reflection at an air gap between two prisms. As a showcase application, we demonstrate the separation of superimposed beams of different diffraction orders directly behind acousto-optic modulators for an operating wavelength of 800\,nm. The wavelength as well as the component size can easily be adapted to meet the requirements of a wide variety of applications. The presented optical element allows to reduce the lengths of beam paths and thus to decrease laser system size and complexity.
\end{abstract}

\section{Introduction}
Obtaining high angular resolution, i.e. separating light beams, which stem from the same source but propagate under slightly different angles, is a common challenge when designing optical systems. Typically, for a given design concept, one can improve the angular resolution by adapting the beam waist or its wavelength and extended propagation distances. However, if waist and wavelength are fixed and system compactness and simplicity are a design goal, alternative approaches need to be considered.

A very common example in this context are acousto-optic modulators (AOMs). AOMs are very well established and widely used devices e.g. for fast switching, frequency shifting and intensity modulation of laser beams and are applied in numerous present-day experiments. Their functional principle relies on light scattering on an acoustic density wave in a crystal. In a typical application, the deflected beam, the undeflected beam or both beams are exploited and thus need to be well separated. As the deflection angle between two diffraction orders is only in the range of $1^\circ$ for an 80\,MHz-AOM in the near-infrared spectrum (IR-A), a beam path of 6\,cm is required to separate beams of waist 0.5\,mm (1/e$^2$ radius) by twice their waist (2.3\,\% overlap).

For applications where compactness of the systems is crucial, like quantum gas experiments in space \cite{Becker2018}, portable optical clocks \cite{Ludlow2018, Grotti2018} or highly complex laser systems with many parallel optical paths like quantum computers \cite{Haffner2008}, beam separation by distance is infeasible. We have therefore developed a highly angular resolving optical beam splitter (``beam separator'') for the specific example of separating undiffracted and first diffraction order beams of an AOM, whose functional principle relies entirely on the angular dependence of total internal reflection.

The wavelength dependencies of our device only stem from the wavelength dependencies of the anti-reflective (AR) coatings, the element can thus be produced for a wide range of wavelengths. Moreover, as the optical element only relies on standard optical elements, it can easily be produced in various sizes and be adapted to different beam diameters.

\section{Operating Principle}
\begin{figure} 
	\centering
	\includegraphics[width=8cm]{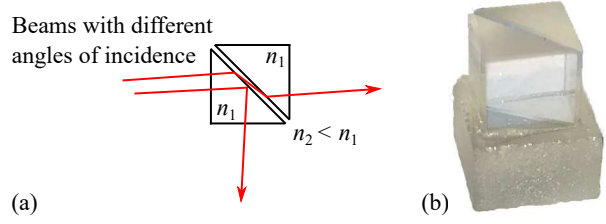}%
	\caption{Two right-angle prisms with index of refraction $n_1$ form a cube with an air gap (index of refraction $n_2<n_1$) at which total internal reflection can occur. When fine-tuned to the critical angle, the beam of one diffraction order is reflected while the other is transmitted through the beam separator.\label{fig:principle}}%
\end{figure}

A schematic top view of the beam separator is depicted in Fig. \ref{fig:principle} (a), a photo in Fig. \ref{fig:principle} (b). Two right-angle prisms with opposite hypotenuses form a cube with an air gap of approximately $125\,\mu$m. Total internal reflection can occur at the interface between the first prism with index of refraction $n_1$ and the air gap with index of refraction $n_2<n_1$ at an incidence angle larger than the critical angle. The critical angle can easily be derived from Snell's law and is given by:
\begin{equation}
\theta_c=\arcsin\left(\frac{n_2}{n_1}\right)~.
\end{equation}

In AOMs, light is diffracted at a sound wave propagating in a crystal. The first order intensity becomes maximum for light incident at the Bragg angle, given by
\begin{equation}
	\theta_B=\frac{\lambda f}{2 v}~,	\label{eq:bragg}
\end{equation}
with $\lambda=\lambda_0/n$ being the wavelength of the light in the crystal, $v$ the acoustic velocity and $f$ the radio frequency (RF). The angle between undiffracted and first diffraction order beams is twice the Bragg angle.

In order to separate the beams of different diffraction orders, the beam separator has to be tuned to the critical angle of total internal reflection so that the beam of one diffraction order is reflected while the other is refracted according to Snell's law. When entering the second prism, this beam is refracted again in the original direction of propagation and therefore transmitted through the optical element.

\section{Implementation}
To build a beam separator, the two prisms made of N-BK7 with 5\,mm long legs (Thorlabs PS909) are glued on a Zerodur base to ensure a fixed width of the air gap. The use of a light-curing adhesive (NOA63) allows for precise alignment of the prisms prior to curing. An optical fiber (without coating) is placed between both prisms as a spacer to ensure the surfaces are parallel. All prism interfaces have an AR coating made in-house for an operating wavelength of 800\,nm. This wavelength has been chosen by way of example and can easily be adapted. At this wavelength, N-BK7 has an index of refraction of 1.5108 leading to a critical angle of $\theta_c=41.45^\circ$.

\begin{figure}
	\centering
	\includegraphics[width=5cm]{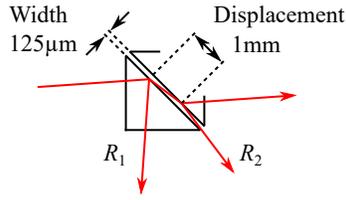}%
	\caption{Beam displacement and unwanted reflections $R_1$ and $R_2$ of the transmitted beam, which are suppressed by AR coatings.\label{fig:reflections}}%
\end{figure}

\begin{figure}
	\centering
	\includegraphics[width=9cm]{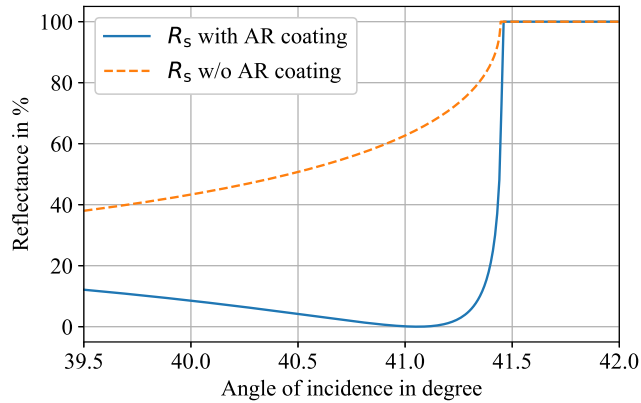}%
	\caption{Calculated reflectance for s-polarized light at a glass ($n=1.5108$) to air interface as a function of the angle of incidence according to Fresnel equations (``$R_\mathrm{s}$ w/o AR coating'' in orange) and with AR coating (blue).\label{fig:reflectance}}%
\end{figure}

While the beam of one diffraction order is completely (``totally'') reflected (independent of the polarization), the transmitted beam is not only refracted but also reflected according to Fresnel equations when leaving the first prism and entering the second (see Fig. \ref{fig:reflections}). We have developed an AR coating with five alternating layers of magnesium fluoride (MgF$_2$, $n=1.4$) and zinc sulfide (ZnS, $n=2.4$) to reduce these unwanted reflections in the vicinity of the critical angle from over 60\,\% to below 1\,\%. The calculated reflectance for s-polarized light as a function of the angle of incidence at a glass-air interface ($R_1$ in Fig. \ref{fig:reflections}) with and without AR coating is depicted in Fig. \ref{fig:reflectance}. The reflectance at the air-glass interface ($R_2$) is equally high because the refraction angle of prism one is equal to the angle of incidence of prism two while the refractive indices are inverted. Therefore, the same coating is applied to both prisms.

The width of the air gap between the two prisms is a trade-off between the influence of beam profile deformation and a potential etalon effect. Due to the large refraction angle, the beam profile is squeezed in one axis which leads to an elliptical profile. Only for small distances between the prisms, there is no need to compensate for this aberration. On the other hand, the distance must be large enough to prevent reflections in itself and thus the occurrence of etalon effects. Since the coupling of light into an optical fiber is very sensitive to changes in the beam profile, we use the fiber coupling efficiency as a figure of merit for the influence of the beam profile squeezing. Etalon effects in turn lead to power fluctuations, so we use power stability as a figure of merit in this case. At the selected width of the air gap of $125\,\mu$m, the beam displacement is approximately 1\,mm (see Fig. \ref{fig:reflections}) for a refraction angle of $41.1^\circ$. In this configuration, we see neither a drop in fiber coupling efficiency nor in power stability compared to beams resolved by a sufficiently long beam path.

\section{Characterization}

\begin{figure}
	\centering
	\includegraphics[width=7.5cm]{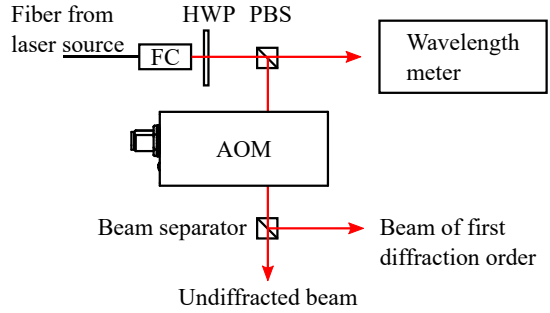}%
	\caption{Schematic setup for characterization measurements of the beam separator. FC: fiber collimator; HWP: half-wave plate; PBS: polarizing beam splitter; AOM: acousto-optic modulator.\label{fig:setup}}%
\end{figure}

A schematic of our characterization measurement setup is depicted in Fig. \ref{fig:setup}. The light of a tunable diode laser (TOPTICA Photonics DL pro) is guided with an optical fiber to the setup. The collimated beam (waist $\approx 400\,\mu$m; divergence angle within the tolerable range of the beam separator, see below) passes a half-wave plate and a polarizing beam splitter for intensity adjustment and polarization cleaning. A wavelength meter (HighFinesse WS6-200) is used to monitor the wavelength in the fraction of the beam transmitted at the polarizing beam splitter. The s-polarized reflected part passes the AOM (A-A MT80-A1.5-IR) with a tellurium dioxide crystal having a material-acoustic mode-velocity of $v=4200$\,m/s and an index of refraction of $n=2.26$ \cite{Young1981}. The AOM is operated at $f=80\,$MHz, the Bragg angle at 800\,nm therefore is [see Eq. (\ref{eq:bragg})] $\theta_B=0.19^\circ$ and the angle between undiffracted and first order beams $0.86^\circ$ outside the AOM according to Snell's law.

\begin{figure}
	\centering
	\includegraphics[width=9cm]{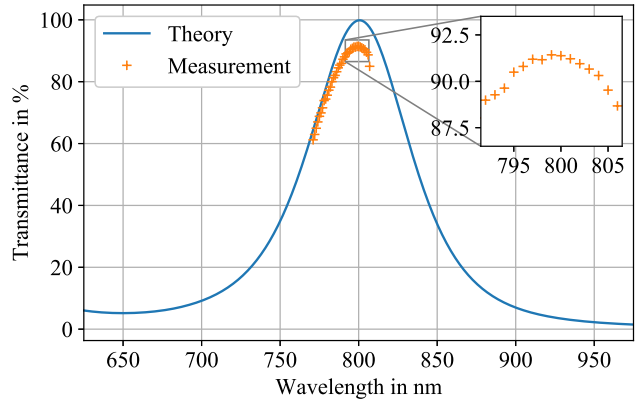}%
	\caption{Calculated transmittance spectrum (blue), given by the squared transmittance of the AR coating for s-polarized light at an angle of incidence of $41.1^\circ$, and measured spectrum (orange). Coating spectrum calculated using OpenFilters \cite{Larouche2008}.\label{fig:transmittance}}%
\end{figure}

For characterization, the beam separator is placed on a tip, tilt, \& rotation stage (Thorlabs TTR001/M) and tuned to the critical angle of total internal reflection. The transmittance spectrum is recorded with respect to a reference power measured between the AOM and the beam separator. The spectrum is shown in Fig. \ref{fig:transmittance} together with the calculated spectrum, given by the squared transmittance of the AR coating for s-polarized light at an angle of incidence of $41.1^\circ$. Despite the already mentioned unwanted reflections plus the losses when entering the first and leaving the second prism, we achieve a maximum transmittance of over 90\,\% (see inset Fig. \ref{fig:transmittance}).

So far we have only discussed beams with long Rayleigh ranges compared to the propagation distances. To reduce the switching times, however, the beam can also be focused into the AOM. As the divergence increases with smaller focus spot size, the question arises which divergence is tolerable by the beam separator and which switching time is therefore achievable. Based on the assumption that both the undiffracted and the first order beam are to be used, both beams should still be separated and no losses should occur. The angle of $0.86^\circ$ between the two beams outside the AOM corresponds to an angle of $0.57^\circ$ inside the beam separator ($n=1.5108$). With the undiffracted beam adjusted to the incidence angle of minimum reflectance at $41.1^\circ$ (see Fig. \ref{fig:reflectance}), this results in an incidence angle of just under $41.7^\circ$ for the beam of the first diffraction order. Thus, a divergence angle of $0.1^\circ$ seems to be achievable without the first-order beam entering the range of transmittance which starts just below $41.5^\circ$. As only the undiffracted beam is entering prism two, there is no comparable criterion for the angular tolerance at this interface. The divergence angle of $0.1^\circ$ allows for a minimum beam waist of approximately 150\,$\mu$m corresponding to a rise time (10\,\%-90\,\%) of 50\,ns compared to 160\,ns for a 1\,mm beam. The rise time can be further reduced if only the undiffracted or the first diffraction order beam is to be used.

\begin{figure}
	\centering
	\includegraphics[width=9cm]{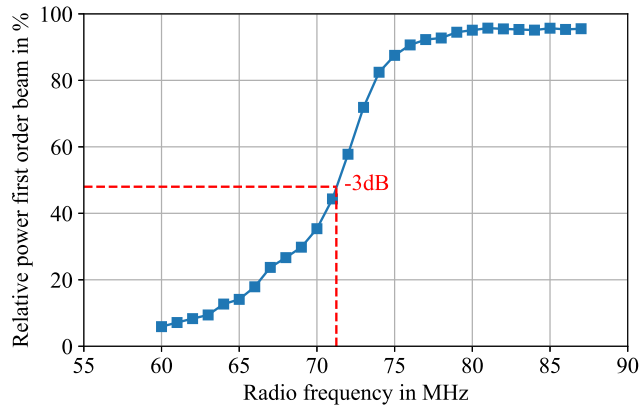}%
	\caption{Power of the first order beam with respect to a reference power measured without beam separator as a function of radio frequency. The beam separator was aligned at 80\,MHz.\label{fig:bandwidth}}%
\end{figure}

Since the deflection angle of the first order beam also correlates with the radio frequency [see Eq. (\ref{eq:bragg})], the angular tolerance of the beam separator also allows to tune the frequency of the AOM e.g. for optical heterodyning. Assuming that the angle between the undiffracted and first order beam outside the AOM is $0.86^\circ$ at 80\,MHz radio frequency, the tolerable angle of $0.1^\circ$ corresponds to 9\,MHz radio frequency. This value is in good agreement with the value obtained by measuring the relative power of the first order beam as a function of radio frequency if the beam separator was aligned at 80\,MHz (see Fig. \ref{fig:bandwidth}). For small radio frequencies, the first order beam moves into the range of transmittance, while for higher frequencies it is always in the range of total internal reflection.

\section{Conclusion}
The optical element presented here separates superimposed beams which differ only by a very small angle. The beams are separated by total internal reflection at an air gap between two prisms. We successfully demonstrated the separation of undiffracted and first diffraction order beams behind AOMs as an example application. The beam separator can be placed directly behind the AOM where the two orders still fully overlap. While the beam of one diffraction order is reflected, the other is refracted twice and therefore transmitted through the optical element. Although the double refraction is theoretically associated with high losses, in practice we achieve transmittances over 90\,\% thanks to our AR coatings. This transmittance requires already accuracies of the coating layer thicknesses in the sub-nanometer range and can only be improved by an even better control of the coating process. The coating also determines the operating wavelength and can easily be adapted as well as the size of the beam separator.

The beam separator enables the development of compact laser system modules and is ideally suited for an implementation in highly stable Zerodur based optical systems \cite{Duncker2014} for space or other field applications. We have already implemented three of the elements in the follow-up system \cite{Anton2017} of the successful sounding rocket mission MAIUS-1 which created the first Bose-Einstein condensate in space \cite{Schkolnik2016, Becker2018}. Moreover, before building the flight hardware, a testbench including the beam separator has been assembled and undergone thermal tests as well as shaker tests imposing vibrations with loads of 8.8\,g$_{\mathrm{RMS}}$. During and after these tests, no damages or malfunctions of the beam separator have occurred. The beam separator is thus well qualified for use in compact and robust laser systems.

\section*{Acknowledgments}
Our work is supported by the German Space Agency DLR with funds provided by the Federal Ministry for Economic Affairs and Energy (BMWi) under grant numbers 50 WP 1433 and 50 WP 1703.




\end{document}